%% file: main.tex
\documentclass[12pt, letterpaper, preprint]{aastex631} \usepackage[T1]{fontenc}
\usepackage{color}
\usepackage{amsmath}
\usepackage{natbib}
\usepackage{ctable}
\usepackage{bm}
\usepackage[normalem]{ulem} % Added by MS for \sout -> not required for final version
\usepackage{xspace}
\usepackage{csvsimple} 

\usepackage{graphicx}
\usepackage{pgfkeys, pgfsys, pgfcalendar}

\let\oldbibliography\thebibliography % killin' me.
\renewcommand{\thebibliography}[1]{%
  \oldbibliography{#1}%
  \setlength{\itemsep}{0pt}%
  \setlength{\parsep}{0pt}%
  \setlength{\parskip}{0pt}%
  \setlength{\bibsep}{0ex}
  \raggedright
}
\newcommand{\given}{\,|\,}

\newcommand{\ourmethod}{{\sc CausalFlow}}

\newcommand{\bitem}{\begin{itemize}}
\newcommand{\eitem}{\end{itemize}}
\newcommand{\beq}{\begin{equation}}
\newcommand{\eeq}{\end{equation}}

\definecolor{orange}{rgb}{1,0.5,0}

\begin{document} \sloppy\sloppypar\frenchspacing 

\title{Exposing Disparities in Flood Adaptation for Equitable Future Interventions}

\newcounter{affilcounter}
\author{Lidia Cano Pecharroman}
\altaffiliation{LCano@mit.edu}
\affil{Department of Urban Studies and Planning, Massachusetts Institute of Technology, 77 Massachusetts Ave, Cambridge, MA 02139, USA}

\author{ChangHoon Hahn}
\affil{Department of Astrophysical Sciences, Princeton University, Peyton Hall, Princeton NJ 08544, USA} 

\begin{abstract}
    As governments race to implement new climate adaptation policies that prepare for 
    more frequent flooding, they must seek policies that are effective for all communities and uphold climate justice. 
    This requires evaluating policies not only on their overall effectiveness but 
    also on whether their benefits are felt across all communities.
    We illustrate the importance of considering such disparities for flood adaptation
    using the FEMA National Flood Insurance Program Community Rating System 
    and its dataset of $\sim$2.5 million flood insurance claims.
    We use \ourmethod, a causal inference method based on deep generative models, to 
    estimate the treatment effect of flood adaptation interventions based on a community's 
    income, diversity, population, flood risk, educational attainment, and precipitation. 
    We find that the program saves communities \$5,000--15,000 per household.
    However, these savings are not evenly spread across communities. 
    For example, for low-income communities savings sharply decline as flood-risk increases in contrast to their high-income counterparts with all else equal.
    Even among low-income communities, there is a gap in savings between predominantly 
    white and non-white communities: 
    savings of predominantly white communities can be higher by more than \$6000 per household. 
    As communities worldwide ramp up efforts to reduce losses inflicted by floods, simply prescribing 
    a series flood adaptation measures is not enough. 
    Programs must provide communities with the necessary technical and economic support to compensate
    for historical patterns of disenfranchisement, racism, and inequality. Future flood adaptation efforts should go beyond reducing losses overall and aim to 
    close existing gaps to equitably support communities in the race for climate adaptation.
\end{abstract}
%\keywords{ keyword1 -- keyword2 -- keyword3 }

% --- intro ---  
\input{intro}
% --- data  ---  
\input{data}
% --- methods ---  
\input{methods}
% --- results ---  
\input{results}

\section*{Acknowledgements}
We would like to thank Peter Melchior, Sebastian Sandoval Olascoaga, Mariana Arcaya, and Lawrence Susskind for their valuable discussions.
LC was supported by the La Caixa Foundation. 
CH was supported by the AI Accelerator program of the Schmidt Futures Foundation.

\appendix
\input{appendix}

\bibliographystyle{mnras}
\bibliography{noah} 
\end{document}

%% file: intro.tex
\section{Introduction} \label{sec:intro}
Flooding constitutes nearly a third of all losses from natural disasters worldwide~\citep{reuter2022}. 
In the US alone, flooding causes more damage than any other severe weather-related  event, with 
annual losses averaging over \$5 billion~\citep{noaa2014}.
These losses are only expected to multiply as climate change raises the sea level 
and increases the frequency of extreme weather events~\citep{rodell2023}.
By the end of the century, rising sea levels and coastal flooding are estimated to cost the 
global economy \$14.2 trillion (a fifth of the global GDP) in damaged assets~\citep{kirezci2020}.   
In response, communities are rapidly enacting flood adaptation measures~\citep{jongman2014, wahl2015, alfieri2016}. 
As these measures have emerged so has evidence of their success~\citep{deegan2007, highfield2017, asche2013,brody2009,davlasheridze2013,kousky2017}.
However, there is still a gap in understanding whether and how the effectiveness
varies across different communities. %y characteristics 

A better understanding of the effectiveness of flood adaptation policies and
their connections to the communities implementing them can ensure that they 
have the intended effect. 
It can also ensure that flood adaptation investments deliver on the desired goals
and effectively allocate limited resources for climate adaptation. 
Furthermore, it can prevent climate interventions from unknowingly replicating 
historical patterns of 
discrimination~\citep{mitchell2015, simpson2022, ranganathan2021}.
This is especially critical in light of recent evidence highlighting patterns 
of inequality in flood preparedness and recovery 
processes in the US~\citep{cutterTemporalSpatialChanges2008, emrich2020,
tateFloodExposureSocial2021, wingInequitablePatternsUS2022, flores2023}. 

We reexamine the effectiveness of flood adaptation interventions 
by evaluating not only {\em whether} they are effective but also {\em for whom}. 
We measure the effectiveness of flood adaption interventions across different 
types of communities using a US wide data set on flood insurance payments
from the National Flood Insurance Program (NFIP) Community Rating System (CRS).
FEMA initiated the CRS in 1990 in order to improve community flood adaptation 
and resilience.
The program is based on a set of prescriptive activities recognized as best practices for flood risk reduction. 
These constitute flood adaptation recommendations that are prevalent in flood planning across the world. 
To join the CRS, communities must implement a series of flood adaptation activities:
\emph{e.g.} floodplain mapping, open space preservation, 
stormwater management activities, or public information and participation programs. 
In exchange, residents of the community receive a discount on their flood 
insurance premium rates.
More than 1,500 out of roughly 20,000 communities in the NFIP 
are currently part of the CRS program~\citep{fema2021}. 

In 2019, FEMA released the NFIP Redacted Claims data set that contains roughly 
2.5 million flood insurance claims. 
It contains claims from communities participating in the CRS as well as  
claims from communities who did not participate in the CRS, but were eligible
for insurance coverage because they complied with 
minimum floodplain regulation requirements. 
Thus, the Redacted Claims data set provides an ideal quasi-experimental setup.
Insurance claims losses, which we use as a proxy for flood loss, can be 
compared between CRS participants and non-participants to quantitatively assess 
the effectiveness of the CRS flood adaptation activities. 

Capitalizing on this quasi-experimental setup, past literature examined whether 
the CRS led to a reduction in flood 
claims~\citep{michel-kerjan2010, davlasheridze2013, highfield2017, kousky2017,gourevitch2023}.
Despite some studies finding the contrary~\citep{asche2013}, the overall consensus
is that flood losses are reduced by the CRS.  
There has been, however, little investigation on whether the program's effectiveness
varies across different types of communities. 
There has also not been any systematic analysis  on the disparities in flood adaptation 
initiatives across communities in the broader international literature.

Past literature has shown that low-income communities respond to risks differently 
to safeguard their livelihoods~\citep{haque2021} and deploy flood-coping mechanisms 
in the absence of flood protection~\citep{brouwer2007}. 
It also highlighted the need to promote pro-poor climate adaptation initiatives and
acknowledge the need need to strengthen low-income household's asset base to improve 
their  adaptation~\citep{mearns2010}. 
Furthermore, some studies have predicted flood losses and vulnerabilities for 
different types of communities~\citep{knighton2020, wing2020, yang2022_floodproperty}.
Nevertheless, no work so far quantifies how the benefits/savings of 
flood adaptation activities are distributed across different communities. 
That is the main goal of this paper. 

Accurately estimating the causal effect of flood adaptation policies and 
its dependence on community characteristics requires modeling their relationship. 
This is especially challenging since the relationship can be highly non-linear, 
complex, and correlated, which violates the assumptions of many standard causal 
inference methods.
In this work, we tackle these challenges using {\sc CausalFlow}, a novel 
method that leverages deep generative models to measure the causal effect in
a data-driven approach. 
With our method, we examine for the first time the effectiveness of the CRS 
as a function of key community characteristics such as population, income, 
diversity, and educational attainment at high-resolution on a zip code level 
over the entire continental US.
We conclusively assess the effectiveness of the CRS flood adaptation 
activities and shed light on who benefits the most from it and under which 
circumstances. 
Our results provide key insight for communities tailoring flood adaptation
interventions. 
It also provides a path forward to re-envisioning flood adaptation in ways that 
can benefit a broader spectrum of communities in the face of climate change. 

%% file: data.tex
\begin{figure*}
    \centering
    \includegraphics[width=0.8\textwidth]{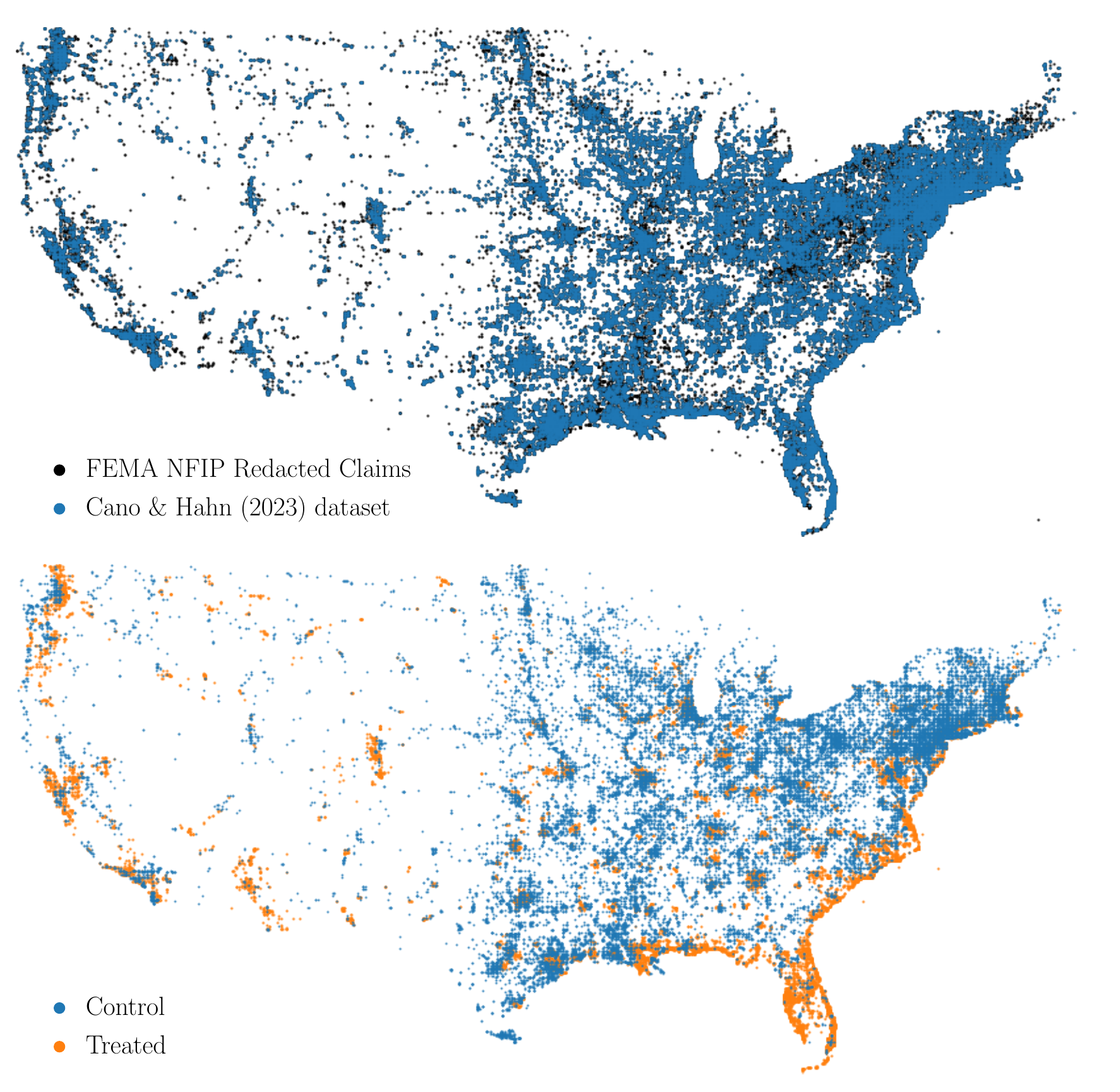}
    \caption{
    {\em Top panel}: We compile our dataset (blue) from the FEMA FIMA NFIP Redacted 
    Claims dataset (black) with additional information on community geo-demographic characteristics
    (precipitation, flood risk, income, population, renter fraction, educational attainment, and diversity fraction) from additional datasets (Section~\ref{sec:data}). 
    {\em Bottom panel}: 
    Communities that participated in the CRS are part of the treated group (orange), while
    the communities that did not are part of control group (blue) 
    }
    \label{fig:map_cate}
    \centering
\end{figure*}

\section{Data} \label{sec:data}
For this work, we compile a dataset based on the FEMA Flood Insurance Mitigation Administration 
NFIP Redacted Claims\footnote{\url{https://www.fema.gov/openfema-data-page/fima-nfip-redacted-claims-v1}}
data with additional information on community characteristics. 
From the NFIP dataset, we use data on CRS participation, the date of flood 
loss, and the total claims paid on building damages and content from the loss. 
We combine all the entries for a zip code and calculate the total claims
paid per policy.

For each zip code, which we refer to as a community, we quantify its flood
risk using scores compiled in the First Street Foundation 
dataset\footnote{\url{https://firststreet.org/data-access/getting-started-with-first-street-data/}}. 
The risk scores are computed based on factors including risk of flooding 
from high-intensity rainfall, overflowing rivers and streams, high tides, 
and coastal storm surges. 
We further supplement the dataset with census data from the 
US Census Bureau American Community Survey\footnote{\url{https://www.census.gov/programs-surveys/acs/news/data-releases.html}}, 
which is compiled in four year intervals: $2008 - 2012$, $2012 - 2016$, and $2016 - 2020$. 
We assign median income and number of residents (population) of the communities based on 
their zip code and date of loss. 
We also calculate the fraction of residents that rent, have a Bachelor's degree or
more advanced degrees, and do not identify as only white. 
We refer to each of the characteristics as the renter fraction, educational attainment, 
and diversity fraction, respectively. 
Lastly, we include average precipitation in millimeters during the month of the flood
event for each community.
This data was extracted by splitting the PRISM climate group data\footnote{\url{https://prism.oregonstate.edu/}} 
compiled by the Northwest Alliance for Computational Science and Engineering using US zip code 
boundaries from 2020.

In total, our dataset includes 14,729 unique communities. 
In the top panel of Figure~\ref{fig:map_cate}, we mark the communities that are 
included in our dataset (blue) from the full NFIP dataset (black) on a map of 
the US. 
Our final dataset is publicly available at XXXXX.

%% file: methods.tex
\section{Methods: {\sc CausalFlow}} \label{sec:methods}
One of the main goals of causal inference is to measure the treatment effect
of a policy, like the CRS. 
For heterogeneous treatments, the effect is quantified using the conditional
average treatment effect (CATE), the ATE as a function of covariates. 
By revealing the dependence of the treatment effect on covariates, CATE 
provides a more detailed understanding of the causal path. 
Given outcome $Y$, covariates $X$, and variable $T$ that indicates the 
control ($T=0$) or treated ($T=1$) groups, CATE is estimated as:
\begin{equation} \label{eq:cate}
    {\rm CATE} = E[~Y \given X, T=1~] - E[~Y \given X,T=0~].
\end{equation}
$E[~Y\given X, T=0,1~]$ represents the expected value of $Y$ given $X$ for 
the control and treated groups, respectively.

Typically, CATE is estimated using either matching or linear regression. 
In matching, samples in the treated group are matched to ones in the control based
on their $X$ values.  
CATE is then estimated by comparing the outcomes of the matched samples. 
Even prevalent methods, such as synthetic control~\citep{abadie2003, abadie2010}
or propensity score
matching~\citep{rosenbaum1983}, match samples based on some finite volume in covariate space, 
which can lead to incorrect estimates of the CATE. 

The other approach is regression, most commonly with linear models~\citep{angrist1990, miguel2004}. 
A model of $Y$ as a linear function of $X$ is fit to the data and then used to
estimate CATE.
In many scenarios, assuming a linear model is incorrect. 
For instance, there is no reason to expect flood losses to depend linearly on its
population, or median household income. 
Furthermore, there is often no a priori knowledge of the functional form that
should be adopted for a model of $Y$.

We can instead estimate the CATE without any of these assumptions.
We rewrite Eq.~\ref{eq:cate} as 
\begin{equation}
    {\rm CATE} = \int Y p(Y|X,T=1)\,{\rm d}Y- \int Y p(Y|X,T=0)\,{\rm d}Y,
\end{equation}
where $p(Y|X,T=1)$ and $p(Y|X,T=0)$ are the conditional probability distribution 
of $Y$ given $X$ for the treated and control groups. 
If we can estimate $p(Y|X,T=1)\approx q_T(Y|X)$ and $p(Y|X,T=0)\approx q_C(Y|X)$
and sample from them, $Y_{T,i}'\sim q_T(Y|X)~{\rm and}~Y_{C,j}'\sim q_C(Y|X)$,
we can estimate CATE using Monte Carlo integration: 
\begin{equation} \label{eq:cate_mc}
    {\rm CATE} =\frac{1}{N_T}\sum\limits_{i=1}^{N_T}Y_{T,i}' - \frac{1}{N_C}\sum\limits_{j=1}^{N_C} Y_{C,j}'.  
\end{equation}

Deep generative models from machine learning (\emph{e.g.} ChatGPT, Dall-E) enable us to 
accurately estimate and sample from $p(Y|X,T=0,1)$.
In this work, %we utilize neural density estimators (NDEs) ---  deep neural networks trained to estimate density distributions. 
we use normalizing flow models~\citep{tabak2010, tabak2013}, 
which use a bijective transformation, $f:z \mapsto x$, that maps a complex target 
distribution, $p(x)$, to a simple base distribution, $\pi(z)$, in our case a 
Gaussian. 
$f$ is defined to be invertible and to have a tractable Jacobian so that target 
distribution can be evaluated from the base distribution: $p(x) =\pi(z) | \det \frac{\partial f}{\partial x}^{-1}|$.
A neural network is used for $f$ to provide an extremely flexible mapping that 
can estimate complex distributions. 
This neural density estimation approach has been used extensively in a variety
of fields spanning neuroscience~\citep[\emph{e.g.}][]{goncalves2020} to
astrophysics~\citep[\emph{e.g.}][]{alsing2019, hahn2022}.

Using normalizing flows\footnote{We use Masked Autoregressive Flow~\citep[MAF;][]{papamakarios2018} models 
implemented by \cite{greenberg2019, tejero-cantero2020}},
we estimate $p(Y|X, T=1) \approx q_{\rm T}(Y\,|\,X)$ and $p(Y|X, T=0) \approx q_{\rm C}(Y|X)$ 
for the treated and control groups separately. 
We describe the training of our normalizing flows $q_{\rm T}$ and $q_{\rm C}$ in 
Appendix~\ref{sec:flows}.
Our outcome, $Y$, is the total insurance claims per policy in dollars. 
We use seven covariates, $X$: precipitation, flood risk, income, population, 
renter fraction, educational attainment, and diversity fraction (Section~\ref{sec:data}). 
The treated group consists of communities participating in the CRS, while the 
control group consists of non-participants.
In the bottom panel of Figure~\ref{fig:map_cate}, we mark the communities in the 
treated (orange) and control (blue) groups of our dataset.

Once trained, we can evaluate CATE at {\em any} given value of the covariates using 
$q_{\rm T}$, $q_{\rm C}$, and Eq.~\ref{eq:cate_mc}, as long as it is within the support of the 
covariates in our data. 
We detail how we ensure this in Appendix~\ref{sec:support}.
\ourmethod~relaxes the strong assumptions made in standard causal inference methods. 
It learns the detailed relationship between $X$ and $Y$ from the data to provide an accurate 
and robust estimate of the treatment effect.

Lastly, we introduce a correction, $\Delta Y$, to the CATE to account for the outreach 
component of the CRS program:
\begin{equation} \label{eq:cate_prime}
    {\rm CATE}' = {\rm CATE} + \Delta Y.
\end{equation}
Communities in the treated group are informed of how to successfully file their claims. 
We estimate in Appendix~\ref{sec:outreach} that the outreach component alone increases 
the total insurance claims per policy by $\Delta Y\sim$ \$9,780. 
Since this increase is not a reflection of any change in flood losses, the CATE 
in Eq.~\ref{eq:cate_mc} underestimates the impact of the CRS. 
Thus, we include $\Delta Y$ and correct for the effect of outreach to more accurately 
quantify the treatment effect on flood loss.
Throughout this work, we refer to ${\rm CATE}'$ as the CRS savings. 

%% file: results.tex
\begin{figure}
    \centering
    \includegraphics[width=\textwidth]{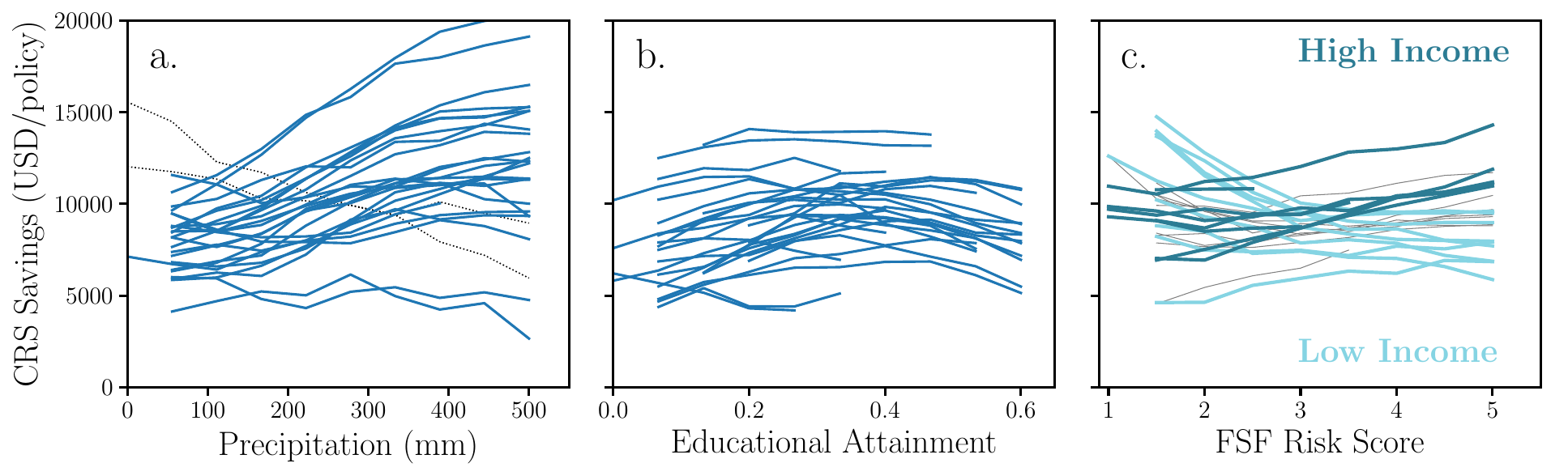}
    \caption{
        Flood loss savings from the CRS program (CATE') for 27 community typologies with
        different population, income, and diversity. 
        In each panel, we vary a single community characteristic, (a) average monthly precipitation, (b) educational attainment, and (c) flood risk score, to highlight 
        its impact on CRS savings. 
        All other characteristics are fixed to fiducial values (Appendix~\ref{sec:typology}).
        The CRS saves policyholders an average of \$5,000--15,000 per policy. 
        However, the efficacy of flood adaption depends significantly on income, 
        population, diversity, precipitation, educational attainment, and flood risk. 
        }\label{fig:cate_multi}
\end{figure}

\section{Results} \label{sec:results}
\subsection{The Effectiveness of Flood Adaptation Measures} \label{sec:success}
With \ourmethod, we can estimate the impact of the CRS flood adaptation activities
on the total insurance claims per policy for any set of community (zip code) 
characteristics.
%,  as long as it is within the support of our dataset. 
In other words, we can measure the ``treatment effect'' of implementing CRS activities and
quantify how much a policyholder saves on average thanks to the program. 
In order to systematize our results, we define 27 distinct community typologies categorized 
by population, income, and diversity (see Appendix~\ref{sec:typology} for details). 
Then, we compute their CATE' for different values of average precipitation, flood risk, 
renter fraction, and educational attainment. 

In Figure~\ref{fig:cate_multi}, we present the CATE' of all 27 community typologies as a
function of (a) average precipitation, (b) educational attainment, and (c) FSF flood risk score. 
In each panel, we vary a single characteristic while keeping all others fixed.
We do this for each typology, represented by a single line. 
This enables us to isolate and examine the effect of a specific characteristic on the CATE'. 
Overall, we find that the CRS saves policyholders an average of
\$5,000 - 15,000 per policy. 
For certain communities, the savings can exceed \$20,000.
Our findings are consistent with previous evidence, which found that the CRS 
led to a 40 percent reduction in losses at the county level~\citep{highfield2017} 
and a \$2.8 -- 5.5 million reduction in damages for a particular flood 
event~\citep{michel-kerjan2010}.  

Beyond estimating the overall savings on flood losses, we can also assess the CRS by 
examining the effect of average precipitation on savings. 
Flood losses are typically worsened by compounded water runoff from higher 
precipitation~\citep[\emph{e.g.}][]{bevacua2019, jang2022, xu2023}.
Yet in Figure~\ref{fig:cate_multi}a, we find that for nearly all of the community  
typologies, the CRS savings increase with higher average precipitation. 
Our results firmly illustrate the CRS program's overall effectiveness in mitigating 
flood losses.

The program's overall success does not paint the full story. 
Our results show that while the CRS is effective, the benefits are not felt
evenly across different communities.
This serves as critical evidence for rethinking and reimagining future flood adaaptation 
policies that aim to equitably reduce flood losses for all communities. 
In the following, we present two lines of action in designing and 
evaluating future strategies. 

\subsection{Tailored Requirements and Resources} \label{sec:tailor}
Community characteristics play a major role in the effectiveness of current flood adaptation measures.
To increase flood savings for {\em all} communities will require tailoring such measures.  
For example, while the CRS is effective at reducing flood losses at higher precipitation,
certain communities go against this trend.
In particular, communities with high population, high diversity, and mid to high income 
(black dotted in Figure~\ref{fig:cate_multi}a), see their savings steeply decrease with 
higher precipitation --- \emph{i.e.} flood adaptation measures are less effective. 
A geospatial analysis of these communities indicates that nearly all of them are in 
urban areas (Figure~\ref{fig:urban}, Appendix~\ref{sec:urban}). 
Urbanized areas have higher percentages of impervious surfaces that increase water runoff 
and can cause or worsen flooding~\citep{pasquierQuantifyingCityScaleImpacts2022a}.  
For these urban communities, flood adaptation programs should require
activities geared towards~\emph{e.g.} decreasing surface imperviousness.  

\begin{figure}
    \centering
    \includegraphics[width=\textwidth]{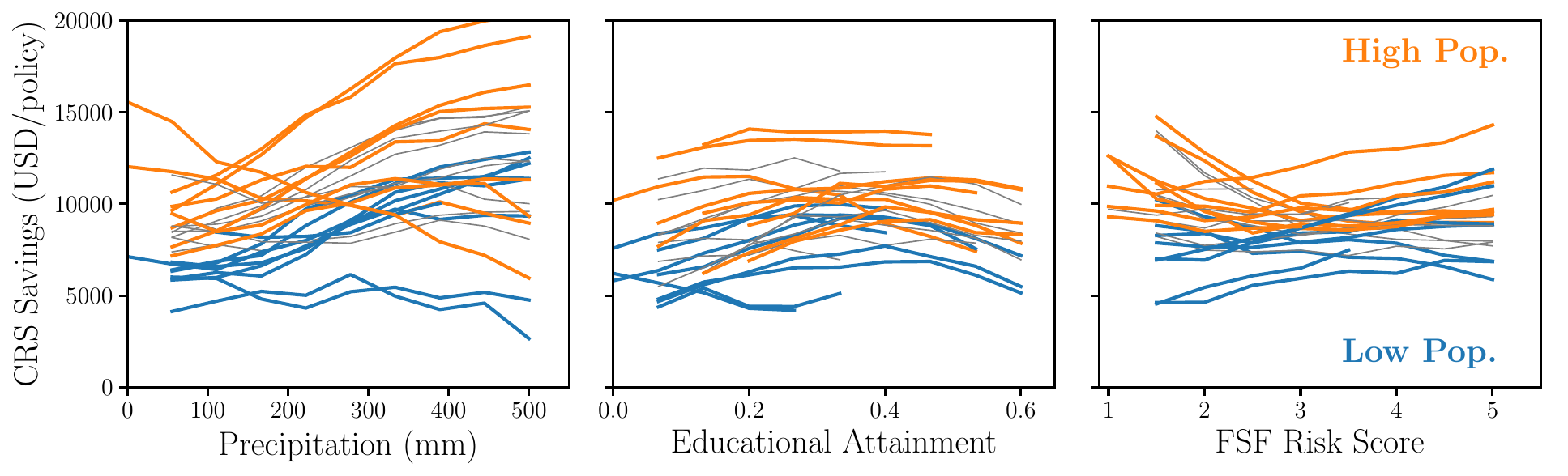}
    \caption{
        The population dependence on the CRS savings. 
        We present the CRS savings for the 27 community typologies, same as in 
        Figure~\ref{fig:cate_multi}, highlighting the high population (orange) and 
        low population (blue) communities. 
        Flood adaptation measures are more effective for high population communities,
        who save $\sim$\$4000 per policy more than low population communities. 
        }\label{fig:cate_pop}
\end{figure}

We also find a significant population dependence. 
In Figure~\ref{fig:cate_pop}, we present the CRS savings for the 27 community typologies 
as a function of precipitation, flood risk score, and educational attainment, 
same as in Figure~\ref{fig:cate_multi}. 
We highlight the communities with a high population in orange and a low population in blue.  
Overall, flood adaptation measures are more effective for high population communities. 
They save $\sim$\$4000 per policy more than less populated communities. 
Combined with previous work, which found that highly populated communities are also more 
likely to adopt CRS activities \citep{asche2013}, our results suggest that flood adaptation 
programs favor populous communities. 

Less populated communities may not have the personpower required to implement the prescribed 
activities,~\emph{e.g.} lacking access to public servants and workers with a wide range
of technical expertise. 
The smaller tax base may also limit the resources available to implement flood adaptation.
In field interviews, we found that the acquisition or relocation of 
flood-prone buildings was limited by the lack of resources available at the local level. 
The implementation of some CRS activities may also require the types of public services 
or resources that are not economically feasible for small communities. 
For instance, the implementation of a flood mapping exercise may prove resource intensive. 
Hence, future programs should provide the necessary technical support and incentivize collaboration among adjacent
communities. 

Finally, we find that the effectiveness of the CRS depends on the communities' educational 
attainment. 
In Figure~\ref{fig:cate_multi}b, we present the CRS savings for the 27 community typologies 
as a function of educational attainment, which we define as the fraction of inhabitants with 
a Bachelor's or higher degree. 
Past evidence already suggests that there is an initial barrier to joining the CRS related
to education, where communities with higher educational attainment are more likely 
to participate~\citep{posey2008, li2012, fan2014}.
Our results show that even after communities join, the effectiveness hinges 
on their education. 
The dependence on education translates into a gap of up to $\sim$\$2,000 per policy in 
savings. 

Out of the 19 creditable activities that communities can implement, many require 
significant capacity building and technical expertise: \emph{e.g.}~the establishment 
of flood warning systems and the building and inspection of levees.
Communities with lower educational attainment may not have the expertise readily 
available and community buy-in may be more difficult. 
In response, future programs should include interventions that reduce the educational/technical
barriers and provide the necessary technical assistance along with tailored community 
outreach resources.  

The trends presented in this section make it clear that future 
flood adaptation programs should tailor their required adaptation measures. 
This is especially the case if program implementation is linked to incentives such as flood insurance price reduction. 
If programs reward a community's adaptation, then not all ``low hanging fruit'' interventions should qualify for the reward. 
Instead, communities should be required to implement the types of flood adaptation interventions 
that are most effective for their needs.
At the same time, future programs  must furnish communities 
with the necessary resources to overcome any financial/technical/educational
barriers in complying with such requirements.
Only then we can expect a just and equitable distribution of the benefits of climate adaptation. 
Next, we illustrate the importance of considering the inequities and disparities of flood adaptation programs.

%2c. Intentionally invest in communities, that may not be able to afford solutions otherwise.
%%%%%%%%%%%%%%%%%%%%%%%%%%%%%%%%%%%%%%%%%%%%%%%%%%%%%%%%%%%%%%%%
\subsection{Climate Justice} \label{sec:justice}
%%%%%%%%%%%%%%%%%%%%%%%%%%%%%%%%%%%%%%%%%%%%%%%%%%%%%%%%%%%%%%%%
An effective and far-reaching flood adaptation strategy requires embedding climate justice at its core. 
Our results show that the effectiveness of a program is tied to economic and racial disparities. 
For example, although flood adaptation is effective overall for low-income communities, 
it is less effective when they are located in high flood risk zones. 
The trend is reversed for affluent communities. 
We highlight this in Figure~\ref{fig:cate_multi}c, where we show CRS savings as a function 
of flood risk for low (light blue) and high-income (dark blue) communities. 
Our results suggests that the most economical CRS activities may only be effective at 
lower risk.
Meanwhile interventions that fare well at higher risk can only be afforded by 
high-income communities. 
Evidently, flood adaptation cannot be left to a community's resources alone, especially since 
average annual flood losses are disproportionately borne by poorer 
communities~\citep{wingInequitablePatternsUS2022}. 

\begin{figure}
    \centering
    \includegraphics[width=\textwidth]{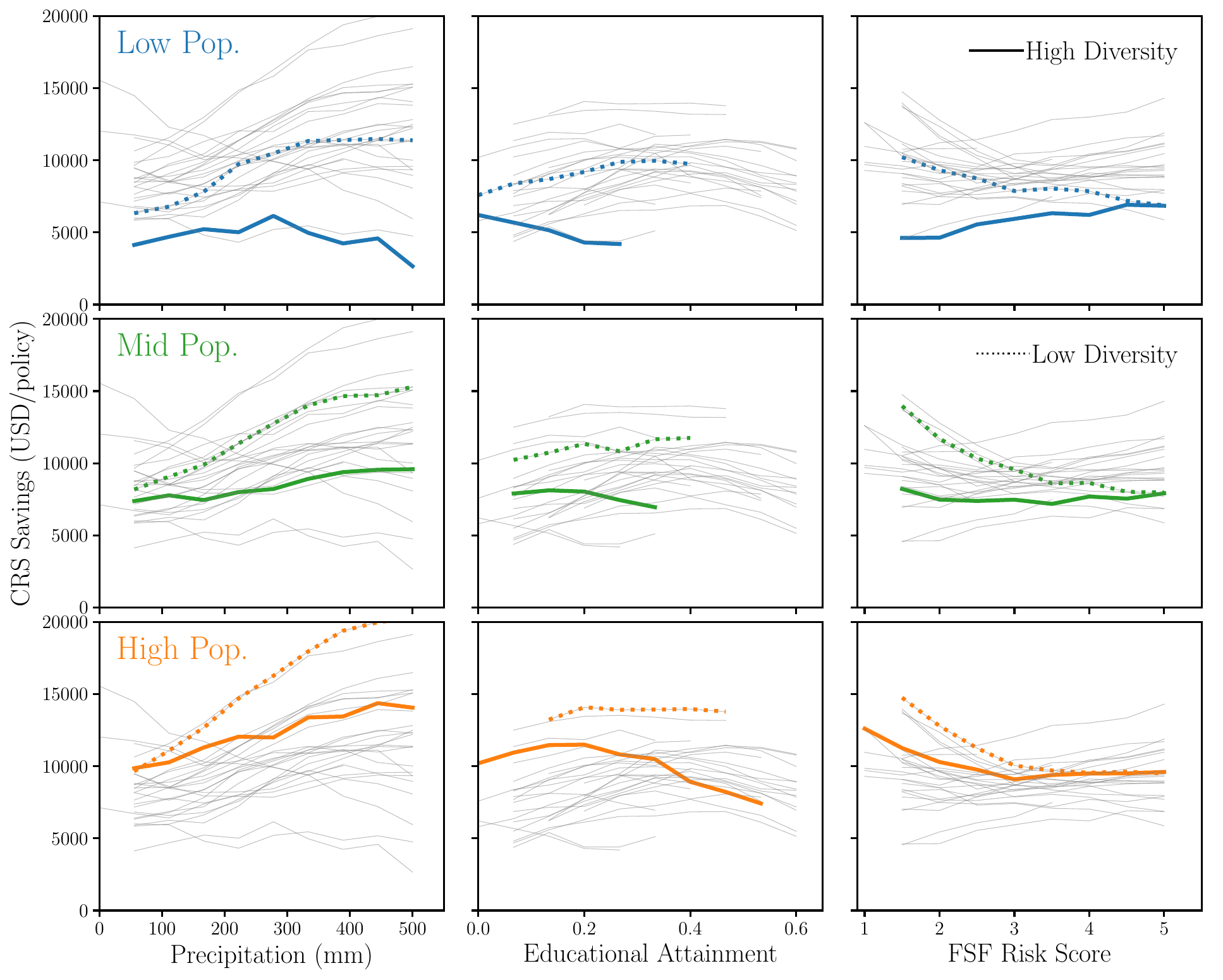}
    \caption{
        CRS savings for low-income communities with high (solid) and low diversity 
        (dotted; predominantly white) as a function of precipitation (left), 
        educational attainment (center), and flood risk (right). 
        We include the other community typologies for reference (gray). 
        Although both communities have the same income level, highly diverse communities
        have significantly lower flood loss savings. 
        The gap between high and low diversity exists for communities with low (blue; top), 
        mid (green; middle), and high (orange; bottom) populations. 
        This diversity gap exceeds \$6,000 per policy in certain cases. 
        }\label{fig:diversity}
\end{figure}

Furthermore, we find that even within low-income communities, there is a systematic gap in the 
CRS savings between diverse and predominantly white (low diversity) communities. 
In Figure~\ref{fig:diversity}, we present the CRS savings for low-income communities with high (solid; predominantly white) and low diversity (dotted; predominantly non-white).
Since the highlighted communities have the same low income, the comparison between the 
solid and dotted lines illustrates the sole effect of diversity. 
Overall, the program is significantly less effective for communities with higher diversity. 
In some cases, the gap exceeds \$6,000 per policy. 
This points to the program’s inability to breach existing patterns of 
discrimination and racial inequality~\citep{mitchell2015, ranganathan2021, simpson2022}.  

This gap is consistent with patterns of inequality found in flood preparedness and 
recovery more broadly~\citep{cutterTemporalSpatialChanges2008, emrich2020,
tateFloodExposureSocial2021, wingInequitablePatternsUS2022}.  
\cite{flores2023} exposed inequalities in the delineation of flood zones in FEMA maps, 
where ``Black and Asian neighborhoods experience disproportionate risk in federally
overlooked pluvial and fluvial flood zones'' (p.1). 
Other works have shown that the type of flood adaptation measures employed correlate with diversity~\citep{sidersVariablesShapingCoastal2020}.
Measures like ``retreat'' correlate with high racial diversity while measures like ``shoreline 
armoring'' correlate with a less diverse population. 
Such differences reflect the real choices that historically disenfranchised communities make based on the 
accessibility of certain types of solutions.

We show that flood adaptation programs can perpetuate institutional racism, through tangible 
effects on savings, or lack thereof. %, that diverse communities can derive from flood mitigation efforts. % that determine on which the community's well-being depends.
Future programs should re-examine the processes, structures, and existing assumptions of flood adaptation 
prescriptions and incentives under the lens of equity, diversity, and inclusion. 
This will be particularly salient as the evidence overwhelmingly shows that the
currently most disadvantaged communities are projected to suffer the most from 
the consequences of climate change. 
Embedding these priorities at the core of future programs will be crucial to close existing gaps 
and support all communities in mitigating flood losses.

\section{Discussion \& Conclusion} \label{sec:discuss}
In this work, we find clear trends between flood loss savings and community characteristics. 
This provides strong evidence that the success of future flood adaptation interventions 
should not only be measured based on whether they reduce losses, but also on whether 
the benefits are equitably distributed. 
Below, we discuss some of the caveats and limitations of our work and outline future research.

First, we note that our results are measured using data from households with 
access to flood insurance.
Hence, our results do not reflect the communities without flood insurance. 
Even though the threshold to access it is relatively low, it is not negligible. 
While some of the communities without insurance are ones that face no significant 
flood risks, they disproportionately include communities without access. 
\cite{kousky2020} found that income of policyholders was higher than non-policyholders, 
which suggests that affordability is a significant concern.
In fact, while a quarter of policyholders are classified as lower income, the fraction is over
a half for non-policyholders~\citep{fema2018}.
Although we cannot quantify flood losses for uninsured communities using 
our dataset, we 
expect that including their losses would only widen the gaps in savings between 
privileged and disenfranchised communities.

We also note that the savings estimates in this work are conservative. 
In calculating the CRS savings, we correct for the outreach component, where communities receive 
information on how to successfully file their claims ($\Delta Y$ in Eq.~\ref{eq:cate_prime}).
This correction is estimated by comparing non-CRS communities to CRS communities that
participated in activities for public information and scored below 10\% on all other 
activities. 
The ideal comparison would be to compare non-participants to CRS communities that only 
participated in public information activities. 
However, this includes very few communities, so we relax this selection.
This likely leads to an underestimate of $\Delta Y$ and, thus, the CRS savings.
Furthermore, our control group consists of communities that did not participate in the CRS but
have access to the NFIP, which requires them to regulate floodplain development. 
Although the requirements are minimal, they may already have a slight effect in reducing 
losses, which would also make our savings estimate conservative.
In subsequent work, we will explore the impact of the outreach component in further detail by 
examining the proportion of successful payments before and after the implementation of the CRS. 
Along these lines, we also find signs that the outreach component could be driving elite capture. 
Savings on flood losses decline for communities with the highest levels of educational attainment
(>50\%; Figure~\ref{fig:cate_multi}c).
Field interviews support the possibility that some households are learning to ``game the system''
in order to refurbish their homes after a flooding event. 
Further research, however, is necessary for a more systematic understanding. 

Our estimate of flood loss savings is also conservative because preventing damage to homes 
mitigates further ripple effects in the livelihoods and well-being of communities. 
Exposure to flood-damaged homes is linked to an increase in mental and health disorders~\citep{graham2019}, death and injury risk,  disease outbreaks~\citep[\emph{e.g.} gastroenteritis;][]{alderman2012},  trauma, anxiety~\citep{walker-springett2017}, and work  disruption~\citep{peek-asa2012}. 
For every dollar saved in flood losses in our results, we can expect a far larger reduction 
in the true loss. 

This study underscores the importance of accounting for complex dependencies when 
evaluating the effectiveness of adaptation efforts and designing future ones. 
The CRS, as a program that prescribes a wide range of flood adaptation interventions, 
serves as an ideal example to showcase the importance of evaluating these efforts from 
a climate justice lens. 
The complex dynamics of climate change require a granular understanding of its effect 
as well as the impact of interventions aimed to combat it.
In this regard, the rapid development of deep generative models presents a unique 
opportunity to move beyond current causal inference approaches.  %parametric model constraints. 
Together with increased investment in data generation, \ourmethod~and similar approaches 
will be capable of addressing previously intractable causal
inference queries that are crucial in designing future policy interventions. 

In summary, this study shows that even though current flood adaptation practices reduce 
flood losses, the savings vary greatly across communities.  
Future adaptation pathways need to consider key community characteristics in 
providing the necessary solutions and resources. 
They must also embed equity priorities at their core to break existing patterns of
inequality and discrimination so that {\em all} communities can  benefit from climate 
adaptation investment into the future.

%% file: appendix.tex
\section{Normalizing Flow Training} \label{sec:flows}
We estimate the CATE of the CRS on total insurance claims per policy using two normalizing flows, 
$q_{\rm T}$ and  $q_{\rm C}$, that estimate the distribution of insurance claims given the covariates 
for the treated and control samples, respectively (Eq.~\ref{eq:cate_mc}).
Below, we describe how we train $q_{\rm T}$ and $q_{\rm C}$ using our data sets from 
Section~\ref{sec:data}. 
We use the same training procedure for both $q_{\rm T}$ and $q_{\rm C}$. 

We begin by splitting  each of the treated and control data sets into training, validation, 
and test sets with a 80/10/10 split. 
We then use the {\sc ADAM} optimizer~\citep{kingma2017} to maximize the total log likelihood 
$\sum_i \log q(Y_i\,|\,X_i)$ over the training set. 
This is equivalent to minimizing the Kullback-Leibler divergence between $q$ and the target 
distribution.
We prevent over-fitting by evaluating the total log likelihood on the validation data at
every epoch and stopping the training when the validation likelihood fails to 
increase after 20 epochs. 

We train $\sim$2000 flows with architectures determined by the \cite{akiba2019optuna} 
hyperparameter optimization.
We select the five flows with the lowest validation losses and construct our final flow as
an equally weighted ensemble of the flows: 
$q_{\phi}(Y\,|\,X) = \sum_{j=1}^5 q_{\phi,j}(Y\,|\,X)/5$. 
Ensembling flows with different initializations and architectures has been shown to improve 
the overall robustness of normalizing flows~\citep[\emph{e.g.}][]{lakshminarayanan2016}. 
After training, we use the test set to verify that $q_{\rm T}$ and $q_{\rm C}$ accurately estimate 
$p(Y|X,T=1)$ and $p(Y|X,T=0)$. 
Specifically, we use simulation-based calibration~\citep{talts2020} and the 
\cite{lemos2023}  coverage test to confirm that both $q_{\rm T}$ and $q_{\rm C}$ are near 
optimal estimates $p(Y\,|\,X,T=1)$ and  $p(Y\,|\,X,T=0)$.

\begin{figure}
    \centering
    \includegraphics[width=0.4\textwidth]{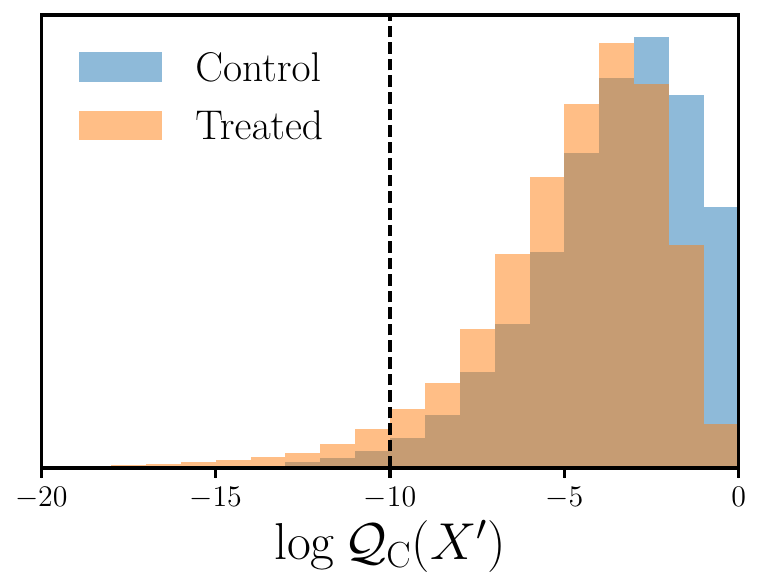}
    \caption{Distribution of $\log \mathcal{Q}_{\rm C}(X')$ evaluated for 
        covariate values in the control (blue) and treated (orange) 
        sample.
        For a given set of covariate values, $X'$ we evaluate 
        $\log \mathcal{Q}_{\rm C}(X')$ and 
        $\log \mathcal{Q}_{\rm T}(X')$. 
        If both are above a conservative threshold (black dashed), we consider 
        it within the support of our treated and control sample. 
        Otherwise, we consider it outside of the support and refrain from 
        evaluating \ourmethod~on $X'$. 
        }\label{fig:support}
\end{figure}

\section{Covariate Support} \label{sec:support}
\ourmethod~can evaluate the CATE at any given value of the covariates within the support of 
the treated and control data, \emph{i.e.} the given covariate is within the covariate 
distribution of the treated and control groups. 
\ourmethod~is based on normalizing flows $q_T$ and $q_C$, 
trained on the treated and control data, respectively.
So evaluating the CATE for any covariate outside their support would mean
evaluating $q_T$ and $q_C$ beyond the covariate distribution they were trained on, 
\emph{i.e.} extrapolation. 
The accuracy of \ourmethod~is not be guaranteed in this regime.
In this work, we avoid this scenario by only evaluating \ourmethod~on 
covariate values classified as being within our treated and control 
distributions.

For the classification, we train two additional normalizing flows, 
$\mathcal{Q}^T(X)$ and $\mathcal{Q}^C(X)$, that estimate the distribution 
of X (community properties) of the treated and control samples. 
We follow roughly the same training procedure outline in Appendix~\ref{sec:flows}; 
however, unlike $q_T$ and $q_C$,
$\mathcal{Q}_{\rm T}$ and $\mathcal{Q}_{\rm C}$ are not conditional distributions. 
For a given coviarate $X'$, we evaluate the log probabilities
$\log \mathcal{Q}_{\rm T}(X')$ and $\log \mathcal{Q}_{\rm C}(X')$. 
If both are above the threshold -10, we classify $X'$ as within the treated
and control support and proceed with evaluating \ourmethod. 
Otherwise, we classify $X'$ as outside of the support. 
The threshold log probability value is empirically determined and conservatively 
set to correspond to the $\sim$1 percentile of the log probability distribution. 
We illustrate this in Figure~\ref{fig:support}, where we present the distribution
of $\log \mathcal{Q}_{\rm C}(X')$ values evaluated on the control (blue) and 
treated (orange) sample. 
We mark our threshold in black dashed. 
Our approach for classifying out-of-distribution covariate values is similar to methods 
for outlier detection~\citep[\emph{e.g.}][]{liang2023, boehm2023}. 

\begin{figure}
    \centering
    \includegraphics[width=0.4\textwidth]{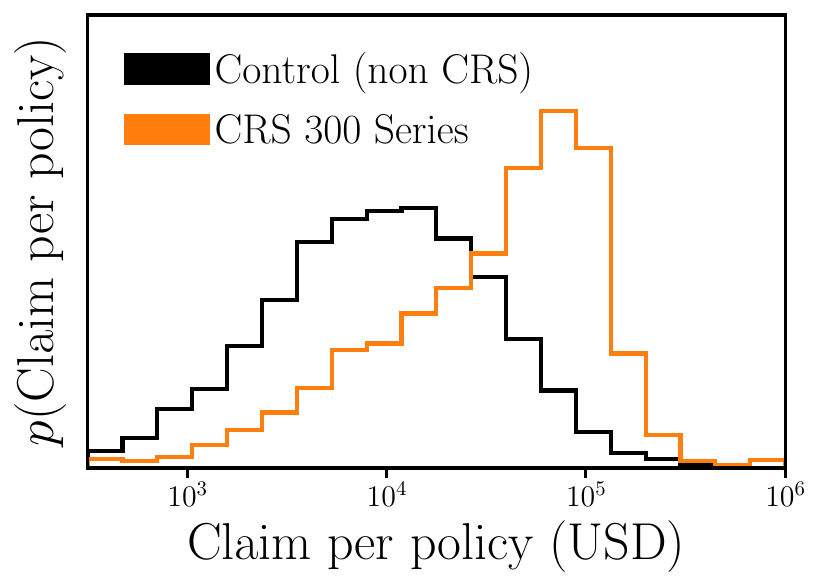}
    \caption{
        The distribution of total insurance claim per policy for non CRS communities (black) 
        and CRS communities that mainly participate in the 300 Series activities (orange). 
        The distribution qualitatively show that the outreach component of the CRS 
        increases claims overall.
        The outreach activities provide guidance for the community on successfully filing 
        claims. 
        Based on a comparison of these two sets of communities in Appendix~\ref{sec:outreach}, 
        we find that the outreach component alone increases claims by $\sim$\$9,780 USD 
        per policy. 
        We correct for this effect in the CATE in Eq.~\ref{eq:cate_prime} to 
        derive CATE', the CRS savings. 
        }\label{fig:outreach}
\end{figure}

\section{Impact of Outreach on CRS Savings}  \label{sec:outreach}
The CRS includes an outreach component where communities are informed how to successfully 
file their flood loss claims.
The outreach itself would not impact actual flood looses. 
Yet, we expect this outreach to increase the total insurance claims per policy for communities 
in the CRS. 
If uncorrected, this would lead to an overall underestimation of the savings from the CRS program. 
Since our primary goal is to accurately measure the impact of the CRS on flood losses, we 
correct for this effect. 
Below, we describe how we estimate the effect of outreach alone. 

FEMA compiles a CRS Communities Credit File data set, which provides detailed information
on the points each CRS community receives for how effectively they implement the various
CRS activities.
Following a request to FEMA, we received access to these scores over the years 2010 to 2020 and 
cross-matched the scores to entries in our NFIP Flood Losses dataset (Section~\ref{sec:data}). 
The activities are categorized into 4 sets of activities: 300, 400, 500, and 600 series. 
They refer to activities for public information, mapping and regulations, flood damage reduction, 
and warning and response, respectively. 
This means that we can isolate the effect of outreach by comparing the insurance claims of 
CRS communities that primarily participate in the 300 activities to our control group, the 
non-CRS communities. 

We first select communities in the CRS that participate in the 300 activities and 
score below 10\% on the 400, 500, and 600 activities. 
These are communities that participate in outreach activities but do not effectively 
implement other flood management activities. 
This ``CRS 300 Series'' sample contains 36 zipcodes with 1955 entries. 
In Figure~\ref{fig:outreach}, we compare the distribution of total insurance claims per 
policy of this sample (orange) to the distribution of our control group (black). 
The comparison reveals that communities participating in outreach have significantly 
higher insurance claims per policy than the control sample. 

To quality the effect of outreach more accurately, we take a similar approach as~\ourmethod. 
For each entry in the ``CRS 300 Series'' sample with covariate values, $X'$, we compute the 
expected insurance claim per policy if the entry was in the control group: 
$E[\,Y\,|\, X',T=0]$ (Eq.~\ref{eq:cate}).
We use the normalizing flow, $q_{\rm C}$, from \ourmethod, with the same procedure 
as Eq.~\ref{eq:cate_mc}. 
We then calculate the difference between the actual insurance claim of the entry and the 
expected control value: $Y'-E[\,Y\,|\, X',T=0]$.
The median value of the effect in the CRS 300 Series sample is \$9,780 USD. 

Our estimate confirms that the outreach component of the CRS indeed significantly increases 
the total insurance claims per policy. 
In principle, if there were more communities in the CRS that only participated in the 300 
activities, we could estimate the effect of outreach as a function of the covariates using
the same approach as \ourmethod. 
However, given our limited sample size, we examine the effect as a function of 
each covariate. 
We find no significant trends. 
Therefore, we use the median value as a fixed correction for the effect of outreach on the 
CRS savings.

\section{Community Typologies} \label{sec:typology}
With \ourmethod, we can examine the effectiveness of flood management activities 
(CATE) as a function of our 7 covariates: income, population, diversity, preciptation, 
flood risk, renter fraction, and educational attainment. 
A 7 dimensional covariate space, however, is challenging to interpret. 
Instead, we systematize our results by examining the CATE for a fixed set of 
27 communities typologies defined by a permutation of income, population, and diversity.

The flood planning literature has shown that \textit{income}, \textit{population}, and \textit{diversity} 
can explain uneven flood damage and loss exposure, as well as differing capabilities 
in preparing for and recovering from floods.
For instance, \cite{tateFloodExposureSocial2021} illustrated that social vulnerability is a crucial indicator of flood exposure. 
Where population, race/ethnicity, and socioeconomic status have been found as the main components of such social vulnerability in relation to natural hazards ~\citep{cutterTemporalSpatialChanges2008}.  
Previous work has also found that community income and race/ethnicity are associated with disproportionate flood impacts and unequal recovery ~\citep{tateFloodExposureSocial2021, wingInequitablePatternsUS2022}.
The contribution of population, income, and race/ethnicity to the communities' vulnerability vary significantly across
counties~\citep{cutterSocialVulnerabilityEnvironmental2003}, so we define our community 
typologies using all of three characteristics.

\begin{table*} 
    \centering
    \begin{tabular}{l|ccc} 
        \hline
         & Low & Mid & High \\
        \hline
        Median Income (USD) & 40,000 & 60,000 & 90,000 \\
        Population & 2,500 & 12,000 & 30,000 \\
        Diversity Fraction & 0.05 & 0.15 & 0.4 \\
        \hline
    \end{tabular} \label{tab:typography}
    \caption{
        Income, population, and diversity fraction values used to 
        define the 27 community typologies for which we examine
        the CRS savings. 
        For the typologies, we use all possible permutations of the
        low, mid, and high values. 
    } 
\end{table*}

For the actual income, population, and diversity values of the community typologies, 
we use values that roughly correspond to the 16, 50, and 84th percentiles of the full data set. 
In Table~\ref{tab:typography}, we list the low, mid, and high 
income, population, and diversity fraction values used in this 
work. 
We take all possible permutations of the values to define 
$3\times3\times3=27$ community typologies.

In addition to income, population and diversity, we also determine the fiducial values of 
the other covariates for the 27 typologies. 
For each typology, we first select communities in our full data set with similar income, 
population and diversity. 
Afterwards set the fiducial covariate value of the typology as the median precipitation, 
flood risk, renter fraction, and educational attainment values of the selected communities.
By taking the median values, we select the covariate valuesto form each typology.

\section{High Population, High Diversity, Mid to High Income} \label{sec:urban}
\begin{figure}
    \centering
    \includegraphics[width=0.5\textwidth]{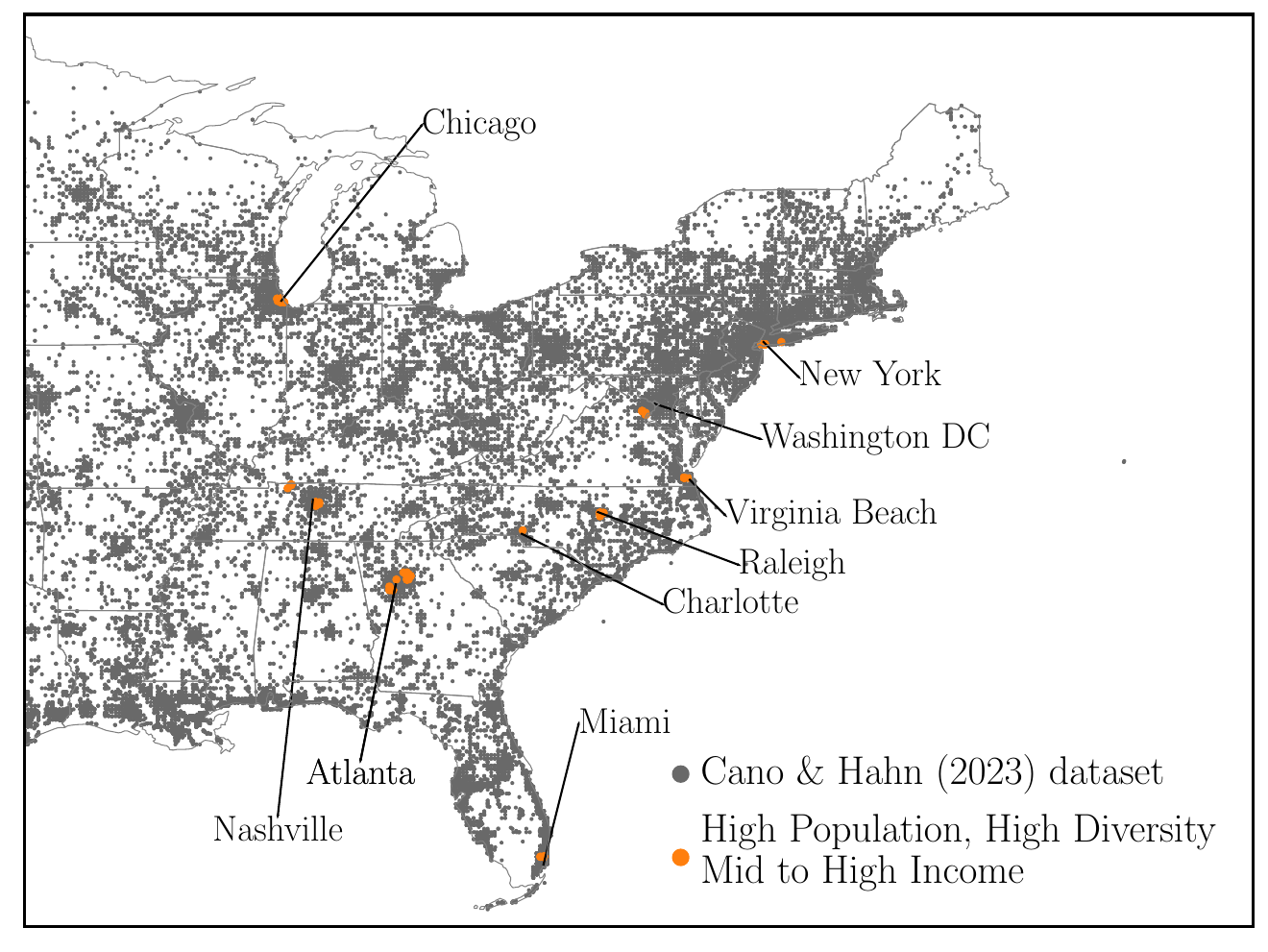}
    \caption{
        We highlight the communities in our data set (gray) that have high population,
        high diveristy, and mid to high income (orange). 
        For reference, we mark the location of major cities near these communities. 
        Compared to other communities in the data set, these communities are
        clearly located in major urban areas.
        }\label{fig:urban}
\end{figure}

In Figure~\ref{fig:urban}, we examine the geospatial distribution of high population,
high diversity, and mid to high income communities (orange) compared to the rest of the
communities in our data set (gray).
The communities are highly localized and adjacent to the major cities that we labeled 
for reference. 
This localization suggests that these communities are in major urban areas.

